\documentclass[journal=NanoLetters,manuscript=article]{achemso}
\setkeys{acs}{keywords = true}
\usepackage[version=3]{mhchem} 
\usepackage[T1]{fontenc}       
\usepackage{multirow}

\author{Mingzhe Yan}
\affiliation
{State Key Laboratory of Low Dimensional Quantum Physics and Department of Physics, Tsinghua University, Beijing 100084, China}
\author{Eryin Wang}
\affiliation
{State Key Laboratory of Low Dimensional Quantum Physics and Department of Physics, Tsinghua University, Beijing 100084, China}
\author{Xue Zhou}
\affiliation
{State Key Laboratory of Low Dimensional Quantum Physics and Department of Physics, Tsinghua University, Beijing 100084, China}
\author{Guangqi Zhang}
\affiliation
{State Key Laboratory of Low Dimensional Quantum Physics and Department of Physics, Tsinghua University, Beijing 100084, China}
\author{Hongyun Zhang}
\affiliation
{State Key Laboratory of Low Dimensional Quantum Physics and Department of Physics, Tsinghua University, Beijing 100084, China}
\author{Kenan Zhang}
\affiliation
{State Key Laboratory of Low Dimensional Quantum Physics and Department of Physics, Tsinghua University, Beijing 100084, China}
\author{Wei Yao}
\affiliation
{State Key Laboratory of Low Dimensional Quantum Physics and Department of Physics, Tsinghua University, Beijing 100084, China}
\author{Shuzhen Yang}
\affiliation
{State Key Laboratory of Low Dimensional Quantum Physics and Department of Physics, Tsinghua University, Beijing 100084, China}
\author{Shilong Wu}
\affiliation
{Hiroshima Synchrotron Radiation Center, Hiroshima University, 2-313 Kagamiyama, Higashi-Hiroshima 739-0046, Japan}
\author{Tomoki Yoshikawa}
\affiliation
{Hiroshima Synchrotron Radiation Center, Hiroshima University, 2-313 Kagamiyama, Higashi-Hiroshima 739-0046, Japan}
\author{Koji Miyamoto}
\affiliation
{Hiroshima Synchrotron Radiation Center, Hiroshima University, 2-313 Kagamiyama, Higashi-Hiroshima 739-0046, Japan}
\author{Taichi Okuda}
\affiliation
{Hiroshima Synchrotron Radiation Center, Hiroshima University, 2-313 Kagamiyama, Higashi-Hiroshima 739-0046, Japan}
\author{Yang Wu}
\affiliation
{Department of Physics and Tsinghua-Foxconn Nanotechnology Research Center, Tsinghua University, Beijing 100084, China}
\author{Pu Yu}
\affiliation
{State Key Laboratory of Low Dimensional Quantum Physics and Department of Physics, Tsinghua University, Beijing 100084, China}
\alsoaffiliation
{Collaborative Innovation Center of Quantum Matter, Beijing, China}
\author{Wenhui Duan}
\affiliation
{State Key Laboratory of Low Dimensional Quantum Physics and Department of Physics, Tsinghua University, Beijing 100084, China}
\alsoaffiliation
{Collaborative Innovation Center of Quantum Matter, Beijing, China}
\author{Shuyun Zhou}
\affiliation
{State Key Laboratory of Low Dimensional Quantum Physics and Department of Physics, Tsinghua University, Beijing 100084, China}
\alsoaffiliation
{Collaborative Innovation Center of Quantum Matter, Beijing, China}
\email{syzhou@mail.tsinghua.edu.cn}
\phone{+86 010 62797928}
\title[An \textsf{achemso} demo]
  {High quality atomically thin PtSe$_2$ films grown by molecular beam epitaxy \footnote{The authors declare no competing financial interest.}}


\begin{document}





\begin{abstract}
  Atomically thin PtSe$_2$ films have attracted extensive research interests for potential applications in high-speed electronics, spintronics and photodetectors. Obtaining high quality, single crystalline thin films with large size is critical. Here we report the first successful layer-by-layer growth of high quality PtSe$_2$ films by molecular beam epitaxy. Atomically thin films from 1 ML to 22 ML have been grown and characterized by low-energy electron diffraction, Raman spectroscopy and X-ray photoemission spectroscopy. Moreover, a systematic thickness dependent study of the electronic structure is revealed by angle-resolved photoemission spectroscopy (ARPES), and helical spin texture is revealed by spin-ARPES. Our work provides new opportunities for growing large size single crystalline films for investigating the physical properties and potential applications of PtSe$_2$.

\end{abstract}


Layered transition metal dichalcogenides (TMDCs) have attracted extensive interests for applications in electronics, optoelectronics and valleytronics due to the strong spin-orbit coupling, sizable band gap and tunability of the electronic structure by quantum confinement effect. \cite{wangqh,ZhangH,xuvalley,YaoWCSR}  In the past decade, this has been witnessed by the significant efforts conducted on the atomically thin MoS$_2$ film. \cite{makMoS,radMoS,makPRL} However, its low mobility has limited applications, for inbstance, in high speed electronics. \cite{MoS2mobility,TMDmobility} Finding thin films of other TMDC with better properties is highly desirable. PtSe$_2$ has emerged as an interesting compound that belongs to TMDC.  Although the bulk crystal is a semimetal, \cite{HHQPRB,KN} monolayer (ML) platinum diselenide (PtSe$_2$) has been revealed to be a semiconductor with a band gap of $\approx$ 1.2 eV. \cite{YWNano}  Importantly, the charge-carrier mobility of PtSe$_2$ has been predicted among the highest in TMDCs \cite{TMDmobility} and has been experimentally shown to be comparable to black phosphorene \cite{BP} yet with the advantage of much improved stability. \cite{ChaiY} This makes PtSe$_2$ a promising candidate for high-speed electronics. Moreover, the hidden helical spin texture with spin-layer locking in monolayer PtSe$_2$  has been recently revealed, \cite{YWSpin} and such spin physics induced by a local Rashba effect has great potential for electric field tunable spintronic devices. \cite{ZungerNP}. In addition, remarkable performance for photocatalytic activity, \cite{YWNano,Voiry,PtSeTiO,compucata} photodetection \cite{photodetector} and quick-response gas sensing \cite{Gassensor} has also been demonstrated. Therefore, PtSe$_2$ is an attractive candidate for a variety of applications. Obtaining high quality PtSe$_2$ films is a critical step toward this goal.

Monolayer PtSe$_2$ film has been first grown by direct selenization of Pt(111) substrate,\cite{YWNano} which is convenient to yield large films up to millimeter size. However, growing such film on metallic Pt substrates hinders the electronics application which instead requires an insulating substrate. The direct selenization method results in a self-terminating monolayer thin film, while thicker films cannot be grown using this method.  Although atomically thin PtSe$_2$ flakes with different thickness can mechanically exfoliated from the bulk crystals, \cite{ChaiY} the sample size is still unsatisfactory and unscalable. Recently, the attempts of growing PtSe$_2$ films by either chemical vapor deposition (CVD) \cite{wangCVD,Raman} or thermally assisted conversion (TAC). \cite{TAC} are reported, however, the polycrystalline nature and lack of atomic-level thickness control are yet to be improved. On the other hand, molecular beam epitaxy (MBE) can provide a better control in terms of growth dynamics and realization of large size, high quality single crystalline films on various substrates with controlled film thickness, \cite{ChoMBE,ArthurMBE} thus providing important material basis for investigating the physical properties and potential applications.
In this work, we report the first layer-by-layer growth of high quality epitaxial PtSe$_2$ thin films on bilayer  graphene/6H-SiC (0001) substrate. The growth process is monitored by reflection high-energy electron diffraction (RHEED) and low-energy electron diffraction (LEED). The high sample quality is revealed by atomic force microscopy (AFM) measurements. We present a systematic study of the vibrational modes and core levels as a function of film thickness by Raman spectroscopy and X-ray photoemission spectroscopy (XPS). Moreover, the band structure measured by angle-resolved photoemission spectroscopy (ARPES) from 1 ML to 22 ML PtSe$_2$ thin films shows the shrinking of band gap as the film thickness increases. Spin-ARPES measurements further  reveal the helical spin texture with spin-layer locking.

PtSe$_2$ crystalizes in stable 1T phase corresponding to the CdI$_2$-type trigonal structure with \textit{P}$\bar{3}$\textit{m}1 space group (No.~162). The building block of PtSe$_2$ contains one Pt atomic layer sandwiched between two Se layers, where Pt atoms are octahedrally coordinated by the Se atoms (Figure 1a). Atomically thin PtSe$_2$ films with varying thickness from 1 ML to 22 ML were grown on bilayer graphene/6H-SiC (0001) substrates using a home built MBE system with a base pressure of 2$\times$10$^{-10}$ Torr. High-purity Se and Pt are evaporated to the substrate with a flux ratio of $\sim$15:1 with the substrate temperature held at 270 $^{\circ}$C. The growth rate is controlled by the Pt flux and excess Se are desorbed from the substrate. The sample thickness is controlled by the growth time. Figure 1b,c show the RHEED patterns before and after growth of the PtSe$_2$ film. RHEED spots from the substrate are completely covered by sharp streaky stripes from the film within the first minutes of growth. The surface morphology of the substrate is revealed by AFM image (Figure 1d), with an average terrace of $\approx$ 200 nm and a step height of $\approx$ 0.75 nm (Figure 1f) which is the typical height of triple SiC bilayer steps.\cite{XueSiC} The overall substrate steps are preserved after growth of the PtSe$_2$ film (Figure 1e), indicating the uniform epitaxy growth along the graphene steps. Compared with samples prepared by CVD or TAC, our MBE films have significantly improved crystallinity, homogeneity and continuity in large scale as well as better thickness control at atomic level. Such high quality single crystalline films provide a unique opportunity for a systematic study of the evolution of the vibrational and electronic properties as a function of sample thickness.

\begin{figure*}
\centering
\includegraphics[width=16.8 cm] {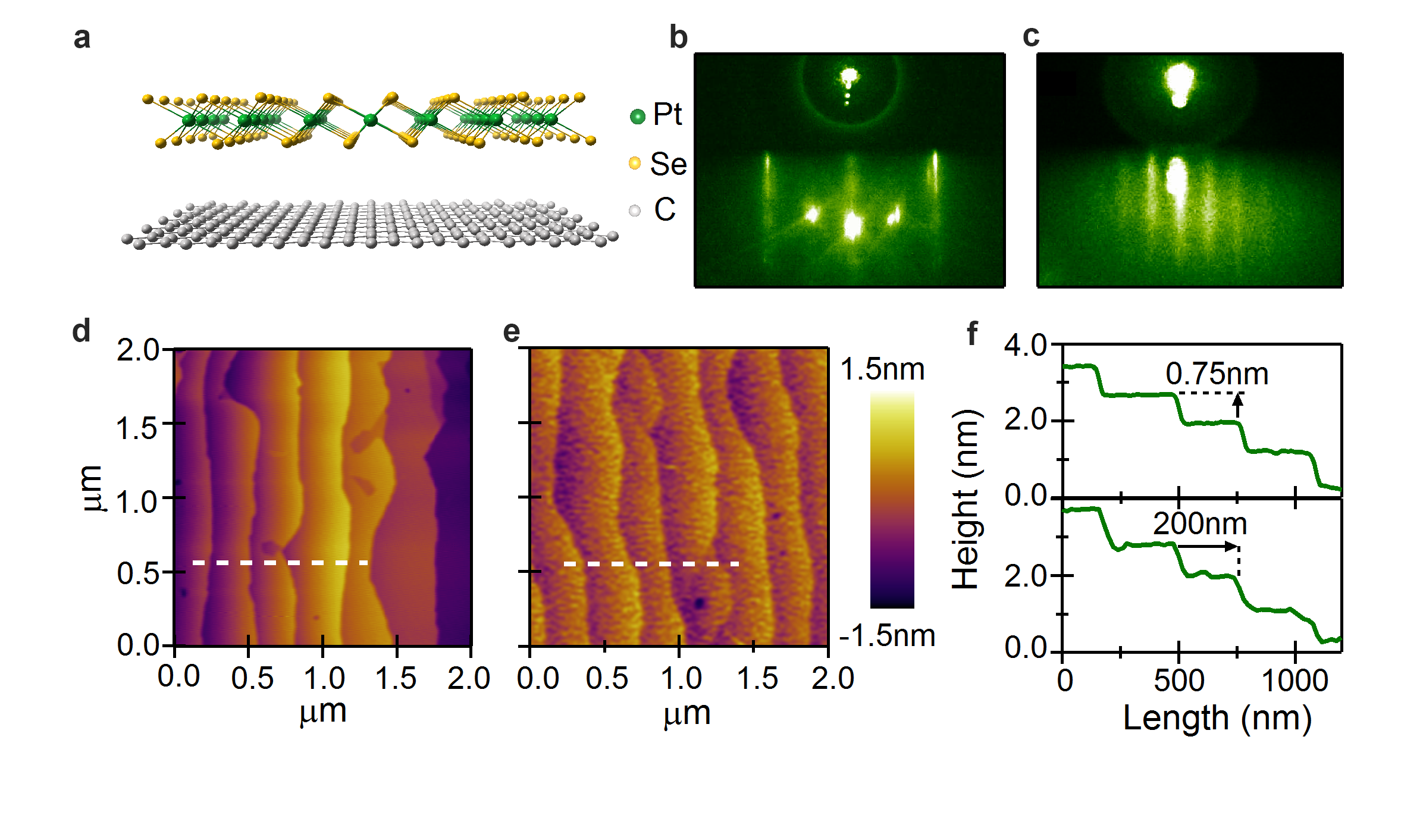}
\label{Figure 1}
\caption{Crystal structure, RHEED pattern and AFM characterization of PtSe$_2$ thin film. (a) Schematic view of PtSe$_2$ grown on graphene substrate. (b)-(c) RHEED patterns of epitaxial bilayer graphene over a 6H-SiC (0001) substrate (b) and after growing a monolayer PtSe$_2$ thin film (c). (d)-(e) Surface morphology images of uniform epitaxial bilayer graphene (d) and 1 ML PtSe$_2$ grown on the substrate (e). (f) The line profile along the dashed line in (d) and (e).}
\end{figure*}

\begin{figure*}
  \centering
  \includegraphics[width=16.8 cm]{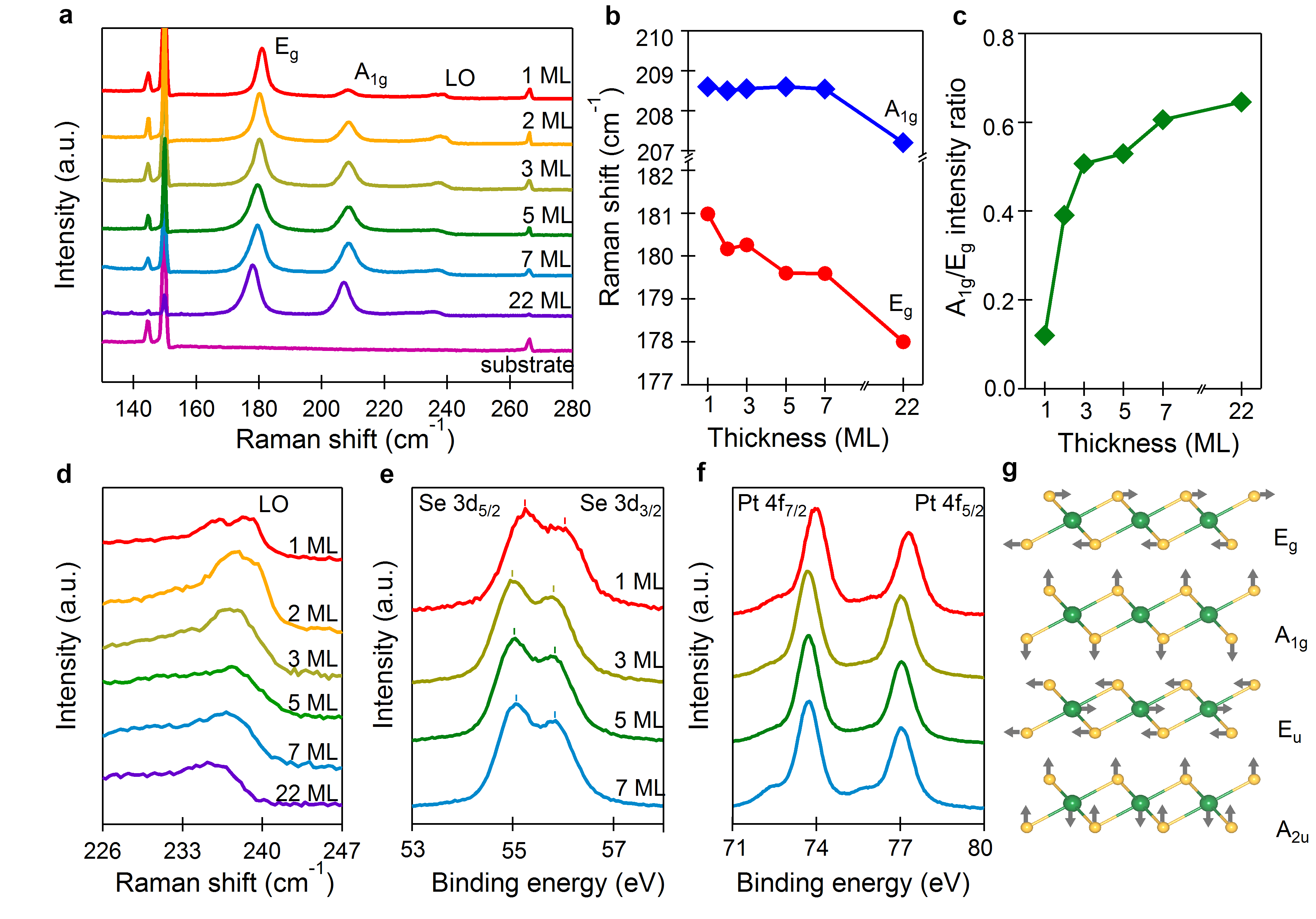}
 \label{Figure 2}
 \caption{Thickness dependent Raman spectra of PtSe$_2$ thin films. (a) Raman spectra of 1, 2, 3, 5, 7, 22 ML PtSe$_2$ films and reference spectrum from the substrate. (b) Evolution of the E$_{g}$ and A$_{1g}$ peak positions with film thickness. (c) Extracted A$_{1g}$/E$_{g}$ intensity ratio in (a).  (d) Zoom-in spectra to see the LO Raman mode in  226-247 cm$^{-1}$. (e)-(f) XPS result of Se 3d and Pt 4f peaks. (g) Schematic views of four Raman-active vibrational modes in PtSe$_2$ layers.}
\end{figure*}

The MBE grown PtSe$_2$ films are characterized by Raman spectroscopy and XPS. Figure 2a shows the Raman spectra of PtSe$_2$ films with varying thicknesses.  The high quality films result in sharper spectra than exfoliated flakes \cite{ChaiY}  and three peaks are clearly identified as E$_g$, A$_{1g}$ and LO. In all PtSe$_2$ films, two prominent peaks at $\sim$180 cm$^{-1}$ and $\sim$208.5 cm$^{-1}$, correspond to the E$_{g}$ and A$_{1g}$ Raman active modes respectively. \cite{Raman,TAC} The E$_{g}$ mode corresponds to an intra-layer in-plane vibration of Se atoms moving in opposite directions and the A$_{1g}$ mode involves the out-of-plane vibration of Se atoms moving away from each other. The E$_{g}$ mode shows a clear blue shift with decreasing film thickness (Figure 2b), while the A$_{1g}$ mode is pinned at 208.5 cm$^{-1}$ at the first few layers thin film. A red shift of A$_{1g}$ peak is also observed in thicker (22 ML) PtSe$_2$ film, which is less significant compared with E$_{g}$ mode. This anomalous behavior may be attributed to stacking-induced structural changes and long-range Coulombic interlayer interactions. \cite{MoSRaman,Raman} Moreover, the peak intensity of the A$_{1g}$ mode relative to that of the E$_{g}$ mode is strongly pronounced from 1 ML to 22 ML (Figure 2c), consistent with the enhanced van der Waals interactions in thicker films. \cite{MoSRaman} The peak at $\sim$240 cm$^{-1}$ is attributed to a longitudinal optical (LO) mode, which is a combination of the out-of-plane (A$_{2u}$) and in-plane (E$_{u}$) vibrations of platinum and selenium atoms respectively, with a similiar origin to those observed in HfS$_{2}$, ZrS$_{2}$ and CdI$_{2}$. \cite{CdI,HfSZrS} The zoom-in spectra in Figure 2d show the strongest LO mode in 1 ML PtSe$_2$ film that splits into two peaks at $\sim$236 cm$^{-1}$ and $\sim$239 cm$^{-1}$. With increasing thickness, such two peaks merge into a broader and weaker peak. Figure 2e-f shows the XPS spectra of Pt and Se.  A  blue shift of Se 3d and Pt 4f core levels by $\approx$ 240 meV (Figure S2 in the supplementary information) suggests the different chemical environment in 1 ML film.  The shift of the core level to higher binding energies suggests that there is likely charge transfer between the substrate and the first PtSe$_2$ layer.

To reveal the evolution of the electronic structure as a function of film thickness, we show in Figure 3 LEED and ARPES data taken from 1 ML film to 22 ML PtSe$_2$ films. Figure 3a-e shows LEED patterns with different film thickness. In 1 ML film (Figure 3a), both the signals from the PtSe$_2$ film (indicated by white arrows) and the substrate (grey arrow) are observed. PtSe$_2$ grows mainly along the orientation of the graphene substrate with some small azimuthal deviations due to the weak coupling between PtSe$_2$ and graphene, showing an arc-like feature in LEED signal. As the growth proceeds (Figure 3b-e), the substrate is completely covered by PtSe$_2$ and the LEED pattern from graphene disappears. ARPES dispersions measured along the $\Gamma$-K high symmetry direction are shown in Figure 3f-j. Our data of 1 ML PtSe$_2$ film (Figure 3f) matches well with previous work on 1 ML PtSe$_2$ film grown by direct selenization method, \cite{YWNano} showing a semiconductor with the top of the valence bands at -1.2 eV. For films thicker than 1 ML, an additional band with an M-shape (indicated by gray arrow in Figure 3g) emerges. This band moves toward the Fermi energy, indicating a reduction of the band gap as predicted. \cite{gapengineer}  The M-shape valence band eventually develops into a three dimensional Dirac cone in the bulk topological Dirac semimetal. \cite{HHQPRB,KN} By increasing the film thickness, the 22 ML PtSe$_2$ film shows effectively the same electronic band structure as the bulk Dirac semimetal except a 350 meV charge transfer from the substrate (Figure S3 in the supplementary information). The ARPES experiments thus provide a direct evidence for the tunable bandgap with varying film thickness.

\begin{figure*}
  \centering
  \includegraphics[width=16.8 cm]{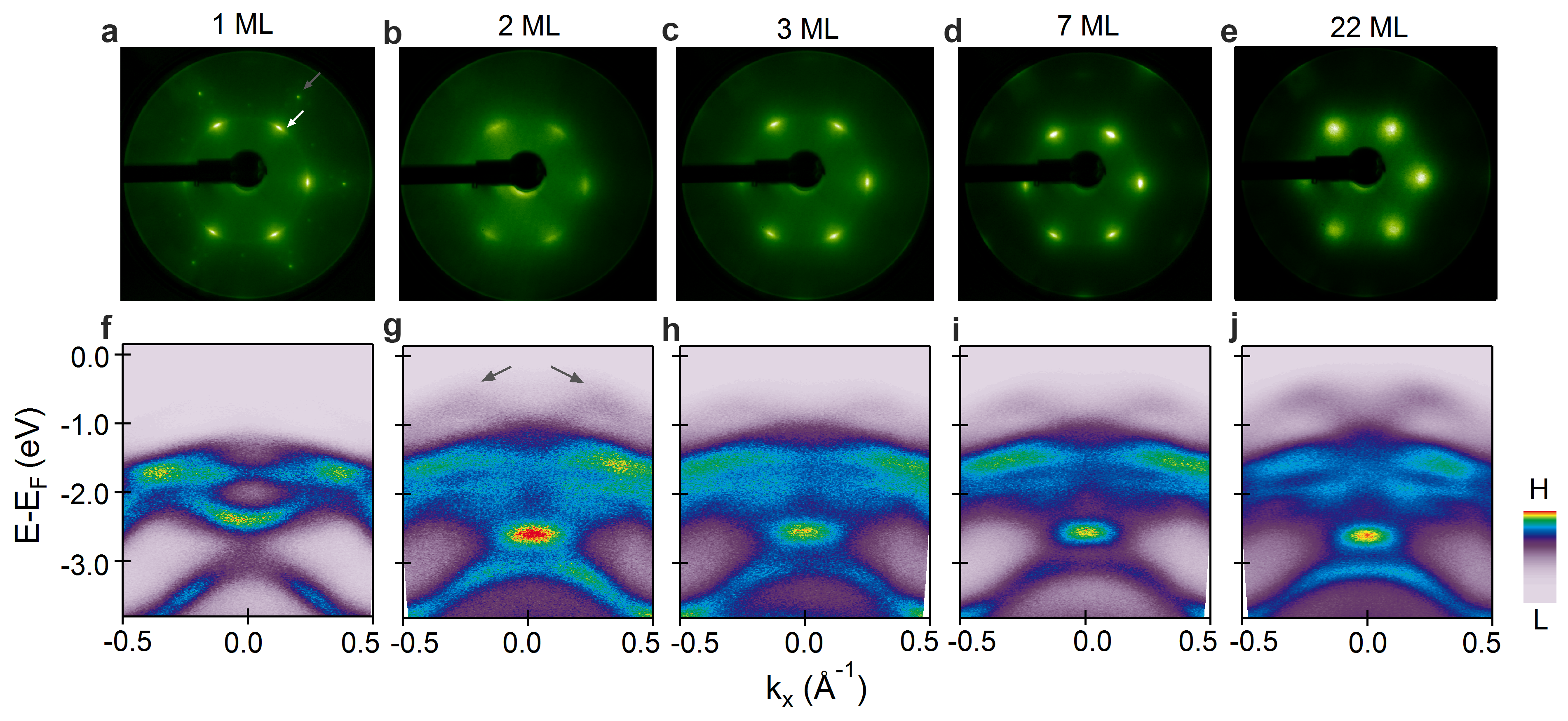}
 \label{Figure 3}
\caption{Electronic structure evolution with thickness in PtSe$_2$. (a)-(e) LEED pattern of 1, 2, 3, 7 and 22 ML PtSe$_2$ thin films. (f)-(j) ARPES spectra of 1, 2, 3, 7 and 22 ML PtSe$_2$ thin films along the $\Gamma$-K direction taken at 21.2 eV.}
\end{figure*}

\begin{figure*}
\centering
\includegraphics[width=13.8cm] {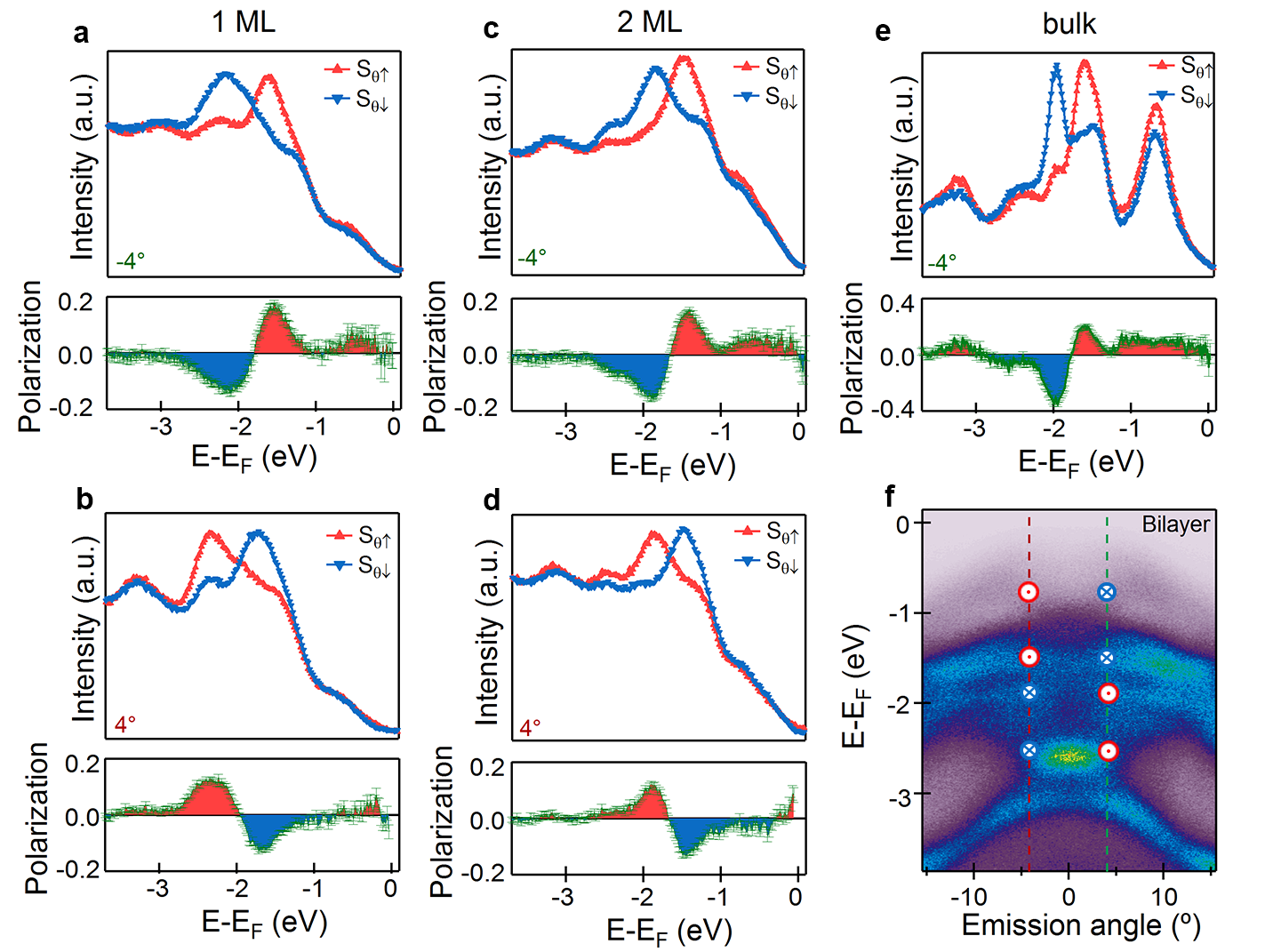}
\label{Figure 4}
\caption{Spin texture of PtSe$_2$ 1 ML and 2 ML films. (a)-(b) Spin-resolved energy distribution curves (EDCs) for the 1 ML PtSe$_2$ in-plane tangential direction at emission angles of $-4^{\circ}$ and $4^{\circ}$, respectively.  (c)-(d) Spin-resolved EDCs for bilayer PtSe$_2$ in-plane tangential direction at emission angles of $-4^{\circ}$ and $4^{\circ}$, respectively. (e) Spin-resolved EDCs for the bulk PtSe$_2$ in-plane tangential direction at emission angles of $-4^{\circ}$. (f) ARPES data of 2 ML PtSe$_2$ film measured at different emission angles.}
\end{figure*}

We further perform spin-ARPES measurements to reveal the spin texture of PtSe$_2$ films. Figure 4a-d shows spin-ARPES measurements along the $\Gamma$-K and $\Gamma$-K$^\prime$ high symmetry directions of both 1 ML and 2 ML films.  Large spin contrast is observed along the tangential direction ($\theta$) at emission angles of $-4^{\circ}$ and $4^{\circ}$, respectively (dahsed line in Figure 4f).  The radial (\textit{\textbf{r}}) and out-of-plane ($\bot$) directions show negligible spin contrast  (Figure S4 in the supplementary information). This is consistent with the  helical spin texture as previously reported in 1 ML PtSe$_2$ films on Pt(111) substrate, \cite{YWSpin}  confirming that it is an intrinsic effect of the PtSe$_2$ film. Similar helical spin texture is also observed in centrosymmetric bulk PtSe$_2$ crystals (Figure 4e), supporting the spin-layer locking mechanism induced by local Rashba effect \cite{Zunger}. Such helical spin texture induced by local Rashba effect makes it useful for electric field tunable spintronics.

To summarize, we have successfully achieved layer-by-layer growth of high quality PtSe$_2$ thin films with controlled thickness using MBE. The samples are characterized by Raman, XPS, LEED and the electronic structure is revealed by APRES. ARPES measurements reveal a distinct tunable bandgap in atomically thin films. The MBE growth can also be extended to insulating substrates. Combined with its unique high charge-carrier mobility and air stability, PtSe$_2$ is a promising candidate for practical applications as new generation electronic devices. Our work reveals the interesting physics in thin PtSe$_2$ films and the MBE growth can in principle be extended to grow large size single crystalline films on a variety of substrates.

\begin{acknowledgement}
This work is supported by the National Natural Science Foundation of China (Grant No. 11334006 and 11427903) and Ministry of Science and Technology of China (Grant No. 2015CB921001 and 2016YFA0301004).
\end{acknowledgement}

\section{METHODS}

Thin film samples of PtSe$_2$ were grown in an ultrahigh vacuum chamber with a base pressure of $2 \times 10^{-10}$ torr. The bilayer graphene substrates were prepared by flash annealing of the 6H-SiC(0001) to 1350 $^{\circ}$C. The growth process was monitored by RHEED and the growth rate was $\sim$30 minutes per monolayer. During the growth process the substrate temperature was kept at 270 $^{\circ}$C. After the growth, the sample was transferred to an ARPES chamber for measurements of the electronic band structure.  ARPES measurements were taken with a Scienta R8000 electron analyzer using UV lamp (21.2 eV) at the temperature of 80 K in a vacuum higher than 1$\times$10$^{-10}$ Torr. XPS spectra of the Pt 4f and Se 3d core-levels were recorded under ultra-high-vacuum conditions better than 1$\times$10$^{-8}$ mbar on a VG Scientific ESCAlab MkII system using Al K$\alpha$ X-rays and an analyzer pass energy of 30 eV.  Raman spectra were measured on a Horiba Raman system with an excitation wavelength of 633 nm and a 1800 lines/mm grating. Spin-ARPES measurements were performed at ESPRESSO endstation of Hiroshima Synchrotron Radiation Center under the temperature of 20 K, using photon energies of 21.2 eV (UV lamp) and 21 eV (Synchrotron radiation). 

%
%
%

\providecommand{\latin}[1]{#1}
\providecommand*\mcitethebibliography{\thebibliography}
\csname @ifundefined\endcsname{endmcitethebibliography}
  {\let\endmcitethebibliography\endthebibliography}{}

\end{document}